\documentclass[prl,amsmath,amssymb,superscriptaddress,twocolumn]{revtex4}
\usepackage{graphicx}

\newcommand{\br}{{\bf r}}

\newcommand{\bk}{{\bf k}}

\newcommand{\bq}{{\bf q}}

\newcommand{\eps}{\epsilon}

\DeclareMathAlphabet{\mathpzc}{OT1}{pzc}{m}{it} 
\begin{document}
\title{Renormalization group approach to 2D Coulomb interacting Dirac fermions with random gauge potential}
\author{Oskar Vafek}
\author{Matthew J. Case}
%\altaffiliation[Also at ]{Department of Physics, FSU}
\affiliation{National High Magnetic Field Laboratory and Department
of Physics, Florida State University, Tallahassee, Florida 32306,
USA}
\address{}
\date{\today}
\begin{abstract}
We argue that massless Dirac particles in two spatial dimensions
with $1/r$ Coulomb repulsion and quenched random gauge field are
described by a manifold of fixed points which can be accessed
perturbatively in disorder and interaction strength, thereby
confirming and extending the results of arXiv:0707.4171. At small
interaction and small randomness, there is an infra-red stable fixed
curve which merges with the strongly interacting infra-red unstable
line at a critical endpoint, along which the dynamical critical
exponent $z=1$.
\end{abstract}

 \maketitle

The properties of two dimensional massless Dirac fermions have
recently sprung back into focus, largely due to the experimental
discovery of the quantum Hall effect in
graphene\cite{Novoselov:2005fk,Zhang:2005uq}, the single layer
graphite. Moreover, the ability to control the density of carriers
by the electrical field effect allows experimental access to the
rich physics of the neutrality point, where in the clean
non-interacting picture, the conduction and the valence bands touch.
It is well known\cite{Gonzales1994} that, at the neutrality point,
the exchange self-energy gives a logarithmic enhancement of the
Fermi velocity $v_F\rightarrow v_F+(e^2/4\epsilon_d)\ln(\Lambda/k)$
where $k$ is a small wavevector near the nodal
point\cite{Gonzales1994} and $\eps_d$ is the dielectric constant of
the medium. Physically, this effect is due to the lack of screening
of the $1/r$ Coulomb interaction, an important consequence of which
is the suppression of the single particle density of states ($N(E)$)
at low energies. This in turn leads to the suppression of the
electronic contribution to the low temperature specific
heat\cite{vafek:prl2007}.

This suppression of $N(E)$ may lure one into the (incorrect)
conclusion that, at $T=0$, the Coulomb interactions turn the clean
system into an electrical insulator. However, the vertex corrections
contribute an {\it exactly} compensating enhancement of the
conductivity\cite{HerbutJuricicVafek2007}, making the system a {\it
metal} with its residual conductivity asymptotically equal to the
non-interacting value $\sigma_0=(\pi/8)e^2/h$ per node.

In this work, we analyze the effects of the unscreened Coulomb
interactions and the quenched random gauge disorder beyond leading
order in the perturbative renormalization group (RG) of Ref.
\cite{HerbutJuricicVafek2007}. Our principle findings, which support
and extend those of Ref. \cite{HerbutJuricicVafek2007} are twofold:
first, in the clean case, there is an unstable fixed point at finite
strength of Coulomb interactions characterized by the dimensionless
ratio $\alpha=e^2/(\hbar \eps_dv_F)$ which represents a quantum
critical point (QCP) separating the semimetal from an excitonic
insulator; and second, the interplay between Coulomb interactions
and disorder induces a downward curvature of the fixed
line\cite{stauber2005,HerbutJuricicVafek2007}, causing it to end at
the clean QCP (see Fig.1).

\begin{figure}[h]
\begin{center}
\begin{tabular}{c}
\includegraphics[width=0.325\textwidth]{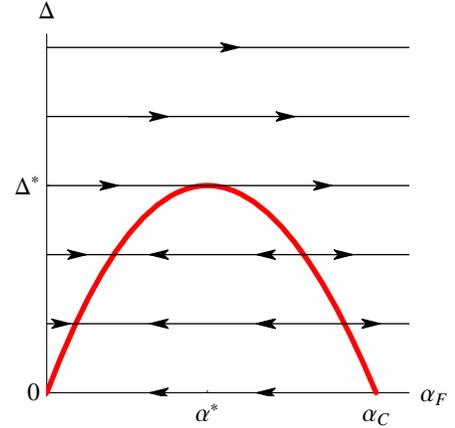}
\end{tabular}
\end{center}
\caption{The renormalization group flow diagram in the (disorder)
$\Delta$ -- (interaction) $\alpha=e^2/(\eps_d v_F)$ plane. There is
a line of stable fixed points at small $\Delta$ and small $\alpha$
which merges with the line of unstable fixed points at the critical
end point. The (clean) unstable fixed point at $\alpha_c$
corresponds to a quantum phase transition into an excitonic
insulator. Above the critical $\Delta^*$, the disordered but
non-interacting fixed line is unstable directly to the insulator.}
\label{fig:RGflows}
\end{figure}

In two dimensions, the Hamiltonian for Coulomb interacting massless
Dirac fermions in the presence of a quenched random gauge field can
be written as
\begin{equation}
\label{H} {\cal H}={\cal H}_0+{\cal H}_{dis}+\hat{V}
\end{equation}
with the free part given by,
\begin{equation}
\label{H0} {\cal H}_0=\sum_{j=1}^N\int{\rm d}^2{\bf
r}v_F\left[\psi_j^\dag({\bf r}){\bf p}\cdot\sigma\psi_j({\bf
r})\right]
\end{equation}
where the operator $\psi_j$ annihilates a two-component Dirac
fermion, ${\bf p}=-i\nabla$, $\sigma_i$ is the $i$th Pauli matrix,
and we have set $\hbar=k_B=1$ for convenience, unless otherwise
stated. $N$ represents the number of fermion species; for single
layer graphene, $N=4$.  The disorder part of the Hamiltonian is
\begin{equation}
\label{Hdis} {\cal H}_{dis}=\sum_{j=1}^N\lambda_jv_F\int{\rm
d}^2{\bf r}\left[a_{\mu}(\br)\psi_j^\dag({\bf
r})\sigma_{\mu}\psi_j({\bf r})\right]
\end{equation}
where $\mu=1,2$ andthe quenched random gauge field is assumed to be
uncorrelated
\begin{eqnarray}
\langle a_{\mu}(\br)\rangle&=&0;\hspace{1cm}\langle
a_{\mu}(\br)a_{\nu}(\br')\rangle=\Delta\delta_{\mu\nu}\delta(\br-\br').
\end{eqnarray}
The charges $\lambda_j=(-1)^j$ vary in sign from node to node as
dictated by the overall time reversal symmetry of the system.
 The Coulomb interaction between Dirac fermions
is given by
\begin{equation}
\label{V} \hat{V}=\frac{1}{2\eps_d}\int d^2{\bf r}d^2{\bf r}'\left[
\delta\hat{n}({\bf r})\frac{e^2}{|{\bf r}-{\bf r}'|}
\delta\hat{n}({\bf r}')\right]
\end{equation}
where $\delta\hat{n}({\bf r})=\sum_j\psi_j^\dag({\bf r})\psi_j({\bf
r})-n_0$ and $e$ is the electronic charge.
The background charge density $n_0$ ensures
overall charge neutrality.

This system can be described by three "coupling constants" -- $e^2$,
$\Delta$ and $v_F$ -- of which $\Delta$ is dimensionless and, in our
units, both $v_F$ and $e^2/\eps_d$ have dimensions of velocity. In
what follows, we will set $\eps_d=1$ and restore it in the final
results by rescaling the charge. Generically, these coupling
constants flow under the renormalization group transformation.
However, the charge $e^2$ does not flow because it is a coefficient
on a non-analytic term in the action, and as detailed elsewhere
$\Delta$ doesn't flow either \cite{HerbutJuricicVafek2007}. The
entire flow of the renormalized coupling constants then comes from
the scale dependence of the Fermi velocity. The RG beta functions
take the form:
\begin{eqnarray}
\frac{de^2}{d\ln\kappa}&=&0\\
\frac{d\Delta}{d\ln\kappa}&=&0\\
\frac{dv_F}{d\ln\kappa}&=&v_F\frac{\Delta}{\pi}-\frac{e^2}{4}
+\mathcal{A}v_F{\Delta}^2+\mathcal{B}e^2\Delta+\mathcal{C}\frac{e^4}{v_F}+\ldots
\label{eq:BetafxnsVF}
\end{eqnarray}
where the ellipses mean terms of cubic order in the double expansion
in small $e^2$ and $\Delta$. The lowest order terms in the expansion
come from the exchange diagrams for disorder and interactions (see
Fig. 2), respectively\cite{HerbutJuricicVafek2007}. The result of
our analysis presented below is the following values of the above
coefficients
\begin{eqnarray}
\mathcal{A}=0, \hspace{0.3cm}
\mathcal{B}=\frac{1}{8\pi},\hspace{0.3cm}
\mathcal{C}=\frac{N}{12}-\frac{103}{96}+\frac{3}{2}\ln2.
\end{eqnarray}
The vanishing coefficient $\mathcal{A}$ agrees with the result of
Ludwig {\it et.al.}\cite{Ludwig:prb94} that, for $e^2=0$, the
dynamical critical exponent $z=1-d\ln v_F/d\ln\kappa=1-\Delta/\pi$
holds to all orders in perturbation theory. The values of
$\mathcal{B}$ and $\mathcal{C}$ are the new results first reported
in this paper.

If we rescale $e^2$ by $v_F$, we can define a dimensionless coupling
constant $\alpha_F=e^2/(\hbar\eps_d v_F)$ which characterizes the
strength of Coulomb interactions. The corresponding flow diagram is
shown in Fig.1. In the clean case, small $\alpha_F$ flows to zero
due to the growth of the Fermi velocity. At $\alpha_F=\alpha_c$,
there is a quantum phase transition into an excitonic insulator,
controlled by the strongly interacting fixed point. Within the above
approximation, for $N=4$, which is appropriate for the single layer
graphene, the unstable fixed point appears at
$\alpha_c=1/4\mathcal{C}\approx 0.833$. As discussed in greater
detail below, since the expansion of the beta function
(\ref{eq:BetafxnsVF}) is carried out in $\alpha_F N\ll 1$, the
semimetal-insulator fixed unstable point appears beyond the strict
validity of the perturbative RG. Nevertheless, for finite $\Delta$,
there is a fixed manifold given by
\begin{eqnarray}
\Delta=\frac{\pi}{4}\frac{\alpha_F(1-4\mathcal{C}\alpha_F)}{1+\pi\mathcal{B}\alpha_F}
\end{eqnarray}
which is shown in Fig.1 by the solid (red) curve. Importantly, at
small $\alpha_F$ and $\Delta$, this manifold represents the line of
stable fixed points \cite{stauber2005,HerbutJuricicVafek2007} which
is asymptotically {\it exact}, and which merges with the line of
unstable fixed points at the critical endpoint
$(\alpha^*,\Delta^*)$.

%\section{Perturbative renormalization group and dimensional regularization}
%
\begin{figure}[t]
\begin{center}
\begin{tabular}{cc}
\includegraphics[width=0.15\textwidth]{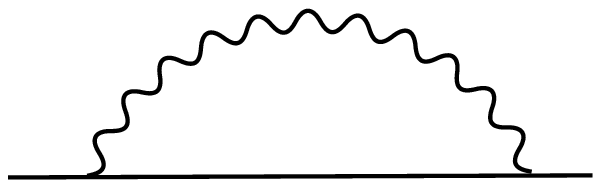}&
\includegraphics[width=0.15\textwidth]{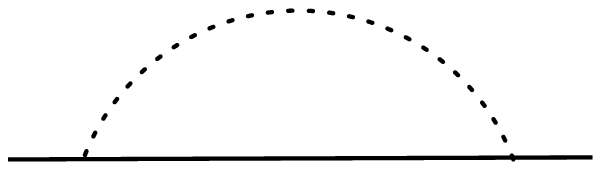}\\
\end{tabular}
\end{center}
\caption{Diagrams contributing to the self energy to the
1$^{st}$order in $\alpha_F$ and $\Delta_A$. As can be readily seen
from the link cluster expansion, the second diagram comes with an
overall $-$ sign.}\label{sigma1st}
\end{figure}
\begin{figure}[t]
\begin{center}
\begin{tabular}{cccccc}
\includegraphics[width=0.15\textwidth]{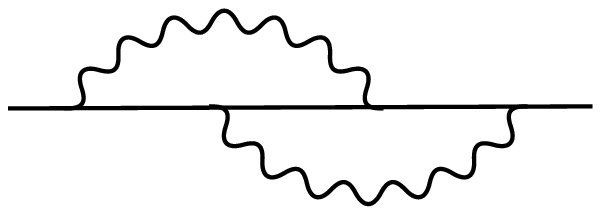}&
\includegraphics[width=0.15\textwidth]{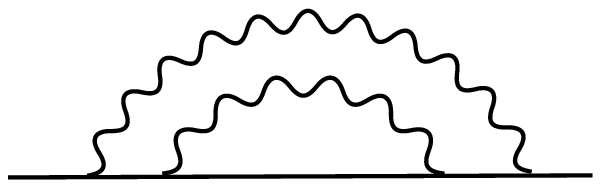}&
\includegraphics[width=0.15\textwidth]{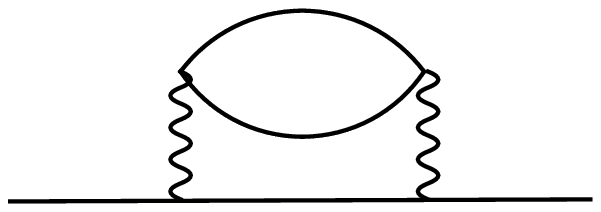}\\
\includegraphics[width=0.15\textwidth]{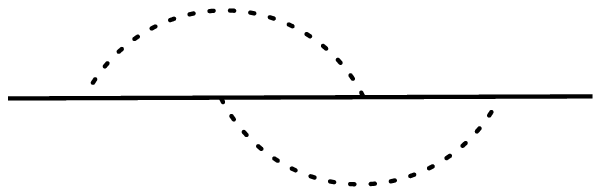}&
\includegraphics[width=0.15\textwidth]{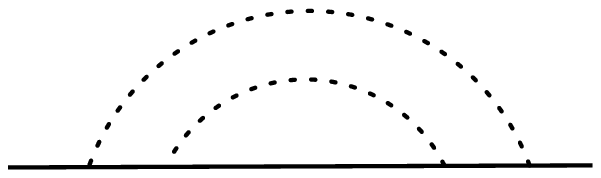}&
\includegraphics[width=0.15\textwidth]{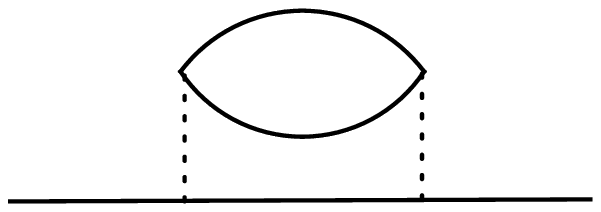}\end{tabular}
\end{center}
\caption{Diagrams contributing to the self energy to the order
$\alpha^2_F$ (first row) and $\Delta^2$ (second
row).}\label{sigma2nd}
\end{figure}
\begin{figure}[t]
\begin{center}
\begin{tabular}{cccccc}
\includegraphics[width=0.15\textwidth]{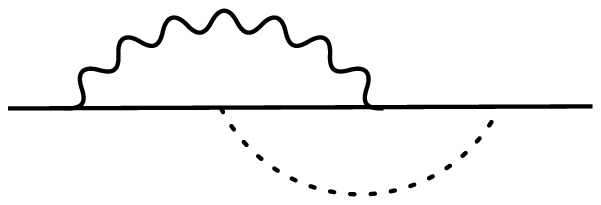}&
\includegraphics[width=0.15\textwidth]{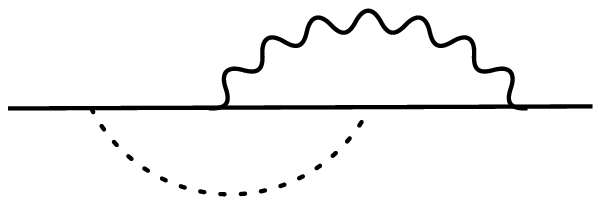}&
\includegraphics[width=0.15\textwidth]{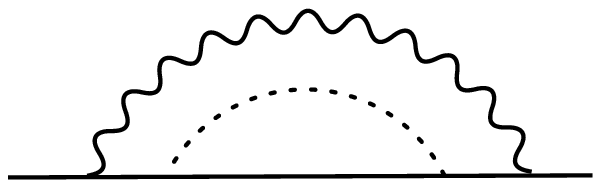}\\
\includegraphics[width=0.15\textwidth]{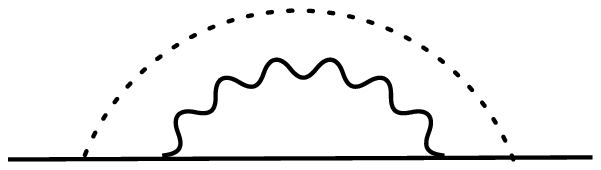}&
\includegraphics[width=0.15\textwidth]{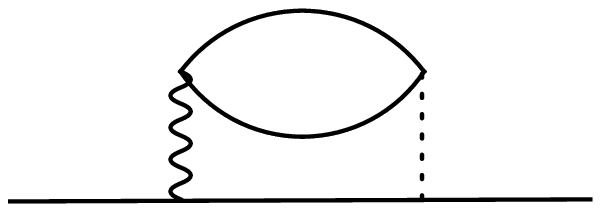}&
\includegraphics[width=0.15\textwidth]{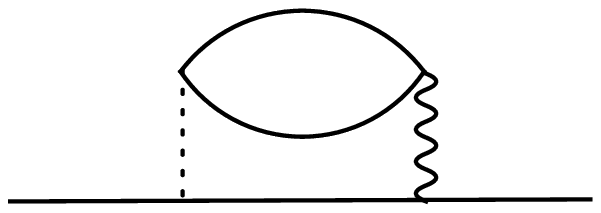}\\
\end{tabular}
\end{center}
\caption{Diagrams contributing to the self energy to the order
$\alpha_F\Delta_A$.}\label{sigma2ndCross}
\end{figure}

Next, we detail the perturbative renormalization group calculation
which we perform in dimensional regularization scheme by
analytically continuing the space integrals to $D=2-\eps$ where
$\eps>0$. This formal device serves as a regulator for various
divergent integrals. The bare Green's function has the form
\cite{Negele:QMP88}
$$
G_0(i\omega,\bk)=(-i\omega+\sigma\cdot\bk)^{-1}=\frac{i\omega+\sigma\cdot\bk}{\omega^2+\bk^2}.
$$
The resulting self-energy matrix, which is defined through the two
point irreducible vertex $\Gamma=\mathcal{G}^{-1}$ as
\begin{eqnarray}\label{eq:bareGamma}
\Gamma_{\bk}(i\omega)=-i\omega+v_F\sigma\cdot\bk+\Sigma_{\bk}(i\omega),
\end{eqnarray}
is calculated at finite external momentum $\bk$ and frequency
$i\omega$. To illustrate the procedure, consider the self-energy to
first order in the coupling constant $e^2$ (the first diagram in
Fig.\ref{sigma1st}):
\begin{eqnarray}
\Sigma_{\bk}^{exch}(i\omega)&=&2\pi
e^2\int\frac{d^D\bq}{(2\pi)^D}\int\frac{d\omega}{2\pi}\frac{1}{|\bk-\bq|}G_0(i\omega,\bq)
\nonumber\\
&=&e^2\sigma\cdot\bk
k^{D-2}\frac{\Gamma[1-\frac{D}{2}]}{(4\pi)^{\frac{D}{2}}}\frac{\Gamma[\frac{D+1}{2}]\Gamma[\frac{D-1}{2}]}{\Gamma[D]}\nonumber\\
&\rightarrow&e^2\sigma\cdot\bk
k^{-\eps}\left(\frac{1}{4\eps}+\frac{1}{8}\ln(64\pi
e^{-\gamma_E})\right).
\end{eqnarray}
In the last line, we have used the analytic properties of the Euler
$\Gamma-$functions and expanded the result near $\eps=0$. The Euler
constant $\gamma_E\approx0.577$. The pole at $D=2$ corresponds to a
logarithmic divergence when a large momentum cutoff is introduced
directly in two dimensions i.e. the Hartree-Fock approximation leads
to the logarithmic enhancement of velocity and
$\Sigma_{\bk}(i\omega)=\frac{e^2}{4}\sigma\cdot\bk \ln\Lambda/k$ is
frequency independent. This logarithmic enhancement of the Dirac
velocity persists to second order in the coupling constant
expansion, in qualitative agreement with
Ref.\onlinecite{mishchenko:prl2007}, as well as the leading order
expansion in large $N$ \cite{Gonzalez:prb99}.

Similarly, the leading order self energy due to the disorder
scattering (second diagram in Fig. \ref{sigma1st}) is
\begin{eqnarray}
&&\Sigma^{dis}_{\bk}(i\omega)=\Delta
\int\frac{d^D\bq}{(2\pi)^D}\sigma_{\mu}G_0(i\omega,\bq)\sigma_{\mu}=-\Delta
i\omega
|\omega|^{D-2}\nonumber\\
&\times&\frac{2\Gamma[1-\frac{D}{2}]}{(4\pi)^{\frac{D}{2}}}
\rightarrow-\Delta i\omega
|\omega|^{-\eps}\left(\frac{1}{\pi\eps}+\frac{1}{2\pi}\ln\left(4\pi
e^{-\gamma_E}\right)\right).
\end{eqnarray}
Again, the pole at $D=2$ corresponds to a logarithmic divergence
of the self-energy.

Before addressing the higher order contributions to the self-energy,
let us define the renormalization conditions. The standard
relationship between the renormalized two point function
$\Gamma^{ren}_{\bk}(i\omega)$ and the bare one (\ref{eq:bareGamma})
is
\begin{eqnarray}\label{eq:Zdef}
\Gamma^{ren}_{\bk}(i\omega)&=&Z\Gamma_{\bk}(i\omega)
\end{eqnarray}
where $Z$ is the wavefunction renormalization
\cite{PeskinSchroeder}. The renormalized coupling constants can now
be defined through the following renormalization conditions
\begin{eqnarray}\label{eq:renConditions1}
\frac{i}{2}\frac{\partial{{\mbox Tr}[\Gamma^{ren}_{\bk}(i\omega)]}}{\partial\omega}\bigg|_{\omega=k=\kappa}&=&1\\
\frac{1}{4}\frac{\partial{{\mbox
Tr}[\sigma_{\mu}\Gamma^{ren}_{\bk}(i\omega)]}}{\partial
k_{\mu}}\bigg|_{\omega=k=\kappa}&=&v^R_F. \label{eq:renConditions2}
\end{eqnarray}
Physically the above equations demand that at the renormalization
scale $\kappa$, the renormalized single particle Green's function
$G^{ren}_{\bk}(i\omega)$ takes the form
$$
{G^{ren}_{\bk}}^{-1}(i\omega)=\left[-i\kappa + v^R_F\sigma\cdot
\kappa\right]^{-1}.
$$

\begin{widetext}
From the above considerations, it is clear that, to find the RG flows,
we need only to find the self energies for $\omega=|\bk|=\kappa$. At
the orders $\Delta^2$, $e^2\Delta$, and $e^4$ we find
\begin{eqnarray}
\Sigma^{(a)}(i\omega,\bk)\bigg|_{\kappa}&=&-i\omega \Delta^2
\frac{\kappa^{-2\eps}}{2\pi^2}\left[
\frac{1}{\eps^2}+\frac{\ln\left(4\pi e^{-\gamma_E}\right)}{\eps}
\right]\\
\Sigma^{(b)}(i\omega,\bk)\bigg|_{\kappa}&=&-i\omega\frac{e^2\Delta}{v_F}\frac{\kappa^{-2\eps}}{4\pi\eps}
+ \sigma\cdot\bk
e^2\Delta\kappa^{-2\eps}\left[\frac{1}{8\pi\eps^2}+\frac{2\ln\left(16\pi
e^{-\gamma_E}\right)+3}{16\pi\eps}\right]\\
\Sigma^{(c)}(i\omega,\bk)\bigg|_{\kappa}&=&
\left[-i\omega-v_F\sigma\cdot\bk\right]
\frac{e^4}{v_F^2}\frac{N}{4}\frac{\kappa^{-2\eps}}{12\eps} -i\omega
\frac{e^4}{v_F^2}\frac{\kappa^{-2\eps}}{4\eps}\left[
-\frac{2}{3}+\ln2
\right]+\sigma\cdot\bk\frac{e^4}{v_F}\frac{\kappa^{-2\eps}}{32\eps}\left[\frac{71}{6}-16\ln2\right]
\end{eqnarray}
\end{widetext}
Thus, to this order in the perturbative expansion, the self energy
matrix in Eq.(\ref{eq:bareGamma}) is
$$
\Sigma_{\bk}(i\omega)=\Sigma^{exch}_{\bk}(i\omega)+\Sigma^{dis}_{\bk}(i\omega)+\Sigma^{a}_{\bk}(i\omega)+
\Sigma^{b}_{\bk}(i\omega)+\Sigma^{c}_{\bk}(i\omega),
$$
and using Eqs.(\ref{eq:Zdef}-\ref{eq:renConditions2}) we find the RG
flow equation for the renormalized Fermi velocity to be
\begin{eqnarray}\label{eq:BetavF}
\frac{dv_F^{R}}{d\ln\kappa}=-\frac{e^2}{4}+v_F^R\frac{\Delta}{\pi}+\frac{e^2\Delta}{8\pi}
+\left(\frac{N}{12}-\frac{103}{96}+\frac{3}{2}\ln2\right)\frac{e^4}{v_F^R}\nonumber.
\end{eqnarray}
This is the result displayed in Eq.(\ref{eq:BetafxnsVF}). It is
apparent from this flow equation that at $\Delta=0$ the Fermi
velocity increases logarithmically provided that the dimensionless
coupling $\alpha<\alpha_c$. Such logarithmic growth implies
suppression of the electronic density of states near the Dirac point
and concomitant suppression of the specific
heat\cite{vafek:prl2007}.

In the case of $\alpha>\alpha_c$, the runaway RG flows can be
interpreted as the flow towards an excitonic insulator. The physical
nature of this insulator depends on the details of the lattice
model. For spinless fermions on a honeycomb lattice, for example, it
would correspond to a state with a spontaneous breaking of
(plaquette or bond centered) inversion symmetry and unequal
population of the two sublattices; for strictly $1/r$ Coulomb
repulsion and for the spinfull case it would correspond to an
"antiferromagnetic" order (but no unit cell doubling) with unequal
spin population of the two sublattices, although details of the
short range part of the repulsive interactions could destabilize it
towards the inversion symmetry broken state\cite{herbutPRL06}.
%At the clean quantum
%critical point, the quasiparticle weight $Z$ vanishes and the
%Green's function becomes
%$$
%G_{\bk}(i\omega)=\frac{i\omega+v^c_F\sigma\cdot\bk}{(\omega^2+{v^c_F}^2\bk^2)^{1-\frac{\eta}{2}}}
%$$ with an exponent
%$$\eta=-\frac{d\ln Z}{d\ln \kappa}=\frac{1}{2}\left(\frac{N}{12}-\frac{2}{3}+\ln2\right)>0,$$
% implying that quasiparticles cannot be
%adiabatically connected to the electrons; the Fermi velocity at the
%QCP is finite and equal to $v_F^c=e^2/(\hbar \eps_d\alpha_c)$.

We conclude with a discussion of the validity of the approximations
employed here. First, the RG was organized perturbatively in both
$\alpha$ and $\Delta$ so the weak coupling portion of the fixed line
in Fig.\ref{fig:RGflows} is rigorously justified. Additionally, the
downward curvature of this line implied by the sign of the terms in
the second order expansion is also rigorous. Diagramatically,
integration of these RG equations corresponds to an infinite
parquet-like resummation of the two leading logarithms at each order
in the pertrubative expansion (see for instance \cite{herbutBook}).

On the other hand, the unstable portion of the fixed line and the
(unstable) clean fixed point appear at a finite value of the
coupling constant at which $\alpha_F N$ is not small. They are,
therefore, beyond the reach of the perturbative RG. Nevertheless,
the very {\em existence} of the clean unstable fixed point is
perhaps on somewhat firmer
footing\cite{KhveshchenkoPRL01,GorbarPRB2002,KhveshchenkoLeal04},
and due to the negative curvature of the IR stable fixed line in the
perturbatively accessible region of the flow diagram, we expect that
the gross topological features of the fixed manifold in
Fig.\ref{fig:RGflows}, if not its quantitative aspects, are valid.

We wish to thank Professors I. Herbut and Z. Tesanovic for useful
discussions and for their critical reading of the manuscript.

\bibliography{RG}
\end{document}